\documentclass[slac_one]{revtex4}
\usepackage{graphicx}
\usepackage{fancyhdr}
\begin{document}


\title{Inclusive Search for the Standard Model Higgs Boson in the 
\boldmath $\mathrm{H} \rightarrow \gamma\gamma$ Channel at the LHC with CMS}


%
\author{M. Gataullin (on behalf of the CMS Collaboration)}
\affiliation{California Institute of Technology,
Department of Physics, M/C 256-48, Pasadena, CA 91125, USA}
\begin{abstract}
 A brief summary of the CMS discovery potential for the Standard Model Higgs boson in
the $\mathrm{H} \rightarrow\gamma\gamma$ channel is presented. We review both a standard
cut-based search and a more optimized analysis that takes advantage of the
wide range of signal/background expectations as function of the possible
selection cuts. As the Higgs discovery in this channel will rely heavily
on performance of the CMS electromagnetic calorimeter, the relevant aspects of its
 design and operation in situ at the LHC are also briefly discussed.
\end{abstract}

\maketitle

\thispagestyle{fancy}

\section{INTRODUCTION}

In the Standard Model of electroweak and strong interactions (SM), 
the fermions acquire mass by interacting with the Higgs field, which is
also assumed to be responsible for the spontaneous breaking of electroweak symmetry. 
At the LHC, the SM Higgs boson is expected to be produced mainly via the gluon fusion although the
vector boson fusion production will also play a significant role. For a Higgs with a mass of 120~GeV,
the corresponding production cross sections are  $\sigma(gg \rightarrow \mathrm{H}) =36.4$~pb 
and $\sigma(qq \rightarrow \mathrm{H}qq) =4.5$~pb.

The LEP2 experiments ruled out a SM Higgs boson lighter than 114 GeV (at 95\% confidence level), 
whereas the current precision electroweak measurements indicate that 
the mass of the SM Higgs should be lower than about 154~GeV (95\%~C.L.)~\cite{bib:lepew}.
Despite its low branching ratio of about 0.002 in this region of interest, the 
 $\mathrm{H} \rightarrow\gamma\gamma$ decay channel will provide an exceptionally clean
final-state topology due to the presence of two energetic photons.
As a consequence, the  $\mathrm{H} \rightarrow\gamma\gamma$ decays are expected to provide 
one of most promising signatures for a discovery with the CMS detector at LHC.


\section{\boldmath $\mathrm{H} \rightarrow\gamma\gamma$ SIGNAL  AT  CMS}

The event selection requires two photon candidates with a $P_T > 35$~GeV each. 
The dominant backgrounds consist of 1) the irreducible background coming from 
the direct diphoton  production  ($\sigma \simeq 150$~pb), and 2) the reducible backgrounds
from
$\mathrm{pp} \rightarrow jet+ \gamma$  ($\sigma \simeq 5\cdot 10^4$~pb) and
$\mathrm{pp} \rightarrow jet+jet$  ($\sigma \simeq 3\cdot 10^7$~pb). 
The signal
rate is thus rather small compared to backgrounds.
To suppress contributions from the reducible background processes, 
both photon candidates are required to be well isolated, 
both in the tracker and in the hadron and electromagnetic calorimeters.
Figure~\ref{fig:1} shows the reconstructed $\gamma\gamma$ invariant mass for  $M_\mathrm{H} = 120$~GeV 
and $\int{\cal L}=1~\mathrm{fb}^{-1}$~\cite{cms:hgg}.

\subsection{CMS Electromagnetic Calorimeter}

\begin{figure*}[t]
\centering
 \includegraphics[width=0.59\linewidth]{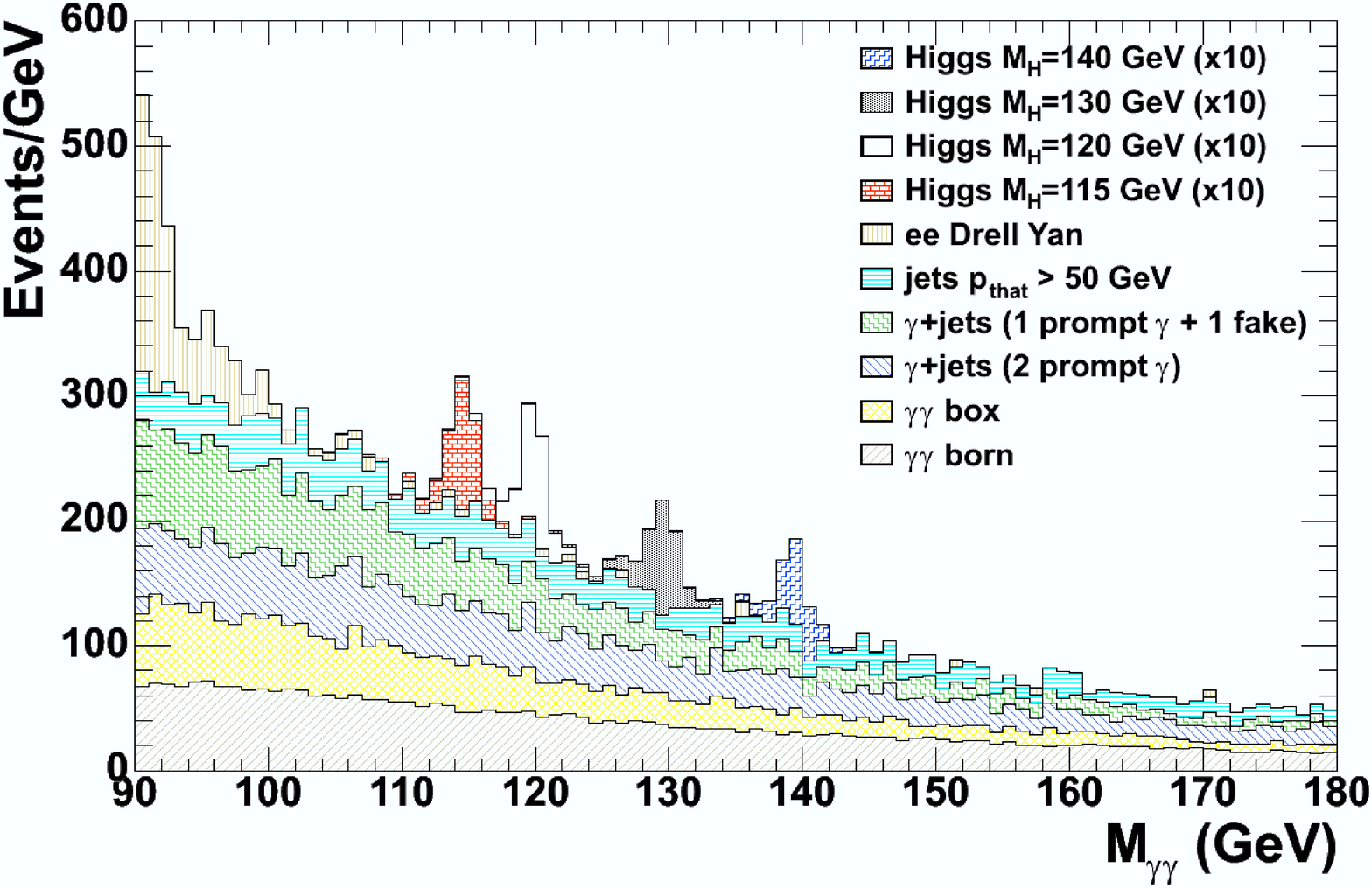}~
 \includegraphics[width=0.405\linewidth]{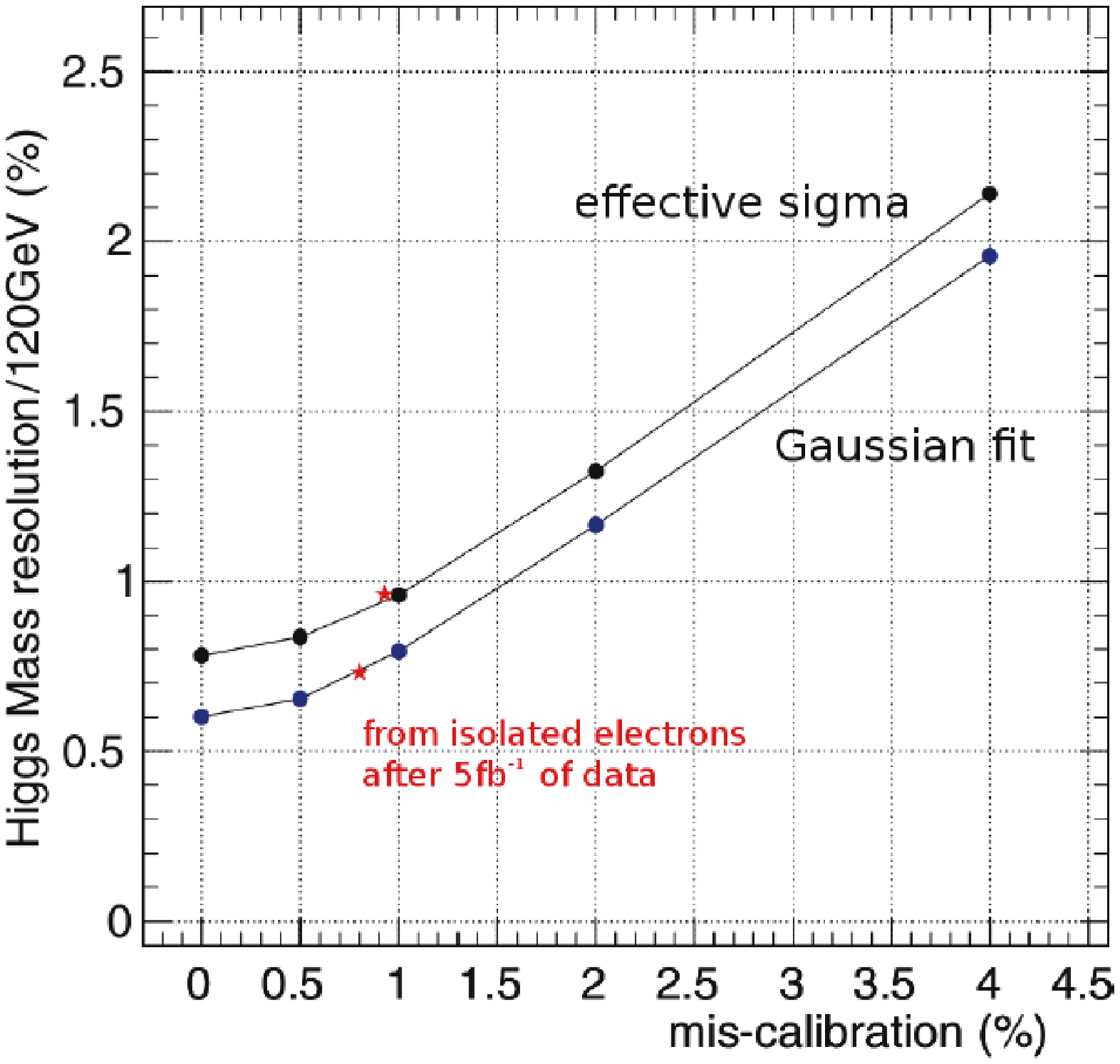}
\caption{Left: 
The diphoton invariant mass spectrum after the selection for the cut-based analysis. Events are normalized
to an integrated luminosity of 1~fb$^ { - 1}$ 
and the Higgs signal, shown for different masses, is scaled by a factor
 10. Right: Relative Higgs mass resolution in the ECAL barrel ($|\eta| < 1.48$) versus initial mis-calibration (no
calibration corrections applied)~\cite{cms:eecal}.}
 \label{fig:1}
\end{figure*}

Made of about 76,000 PbWO$_4$ crystals, the CMS ECAL is one of the best calorimeters in high-energy 
physics,
designed to give CMS a superior discovery potential for the Higgs in the critical low-mass range. 
The target energy resolution of $0.5\%$ for electrons and photons with an energy
above about $50$~GeV has been successfully achieved in several test beam studies.
As shown in Figure~\ref{fig:1}, precise, in situ calibration and monitoring of the 
ECAL response will be crucial for a clean reconstruction of the  $\mathrm{H} \rightarrow\gamma\gamma$ signal.
 This task will be accomplished using a dedicated laser monitoring system and calibrations
 with $\pi^0 \rightarrow \gamma\gamma$, $\mathrm{Z} \rightarrow \mathrm{e}^+\mathrm{e}^-$, and 
$\mathrm{W}\rightarrow \mathrm{e}\nu$ decays and the ``$\phi$-invariance'' method~\cite{cms:calib}.

\section{DISCOVERY POTENTIAL}
\begin{figure*}[t]
\centering
 \includegraphics[width=0.545\linewidth]{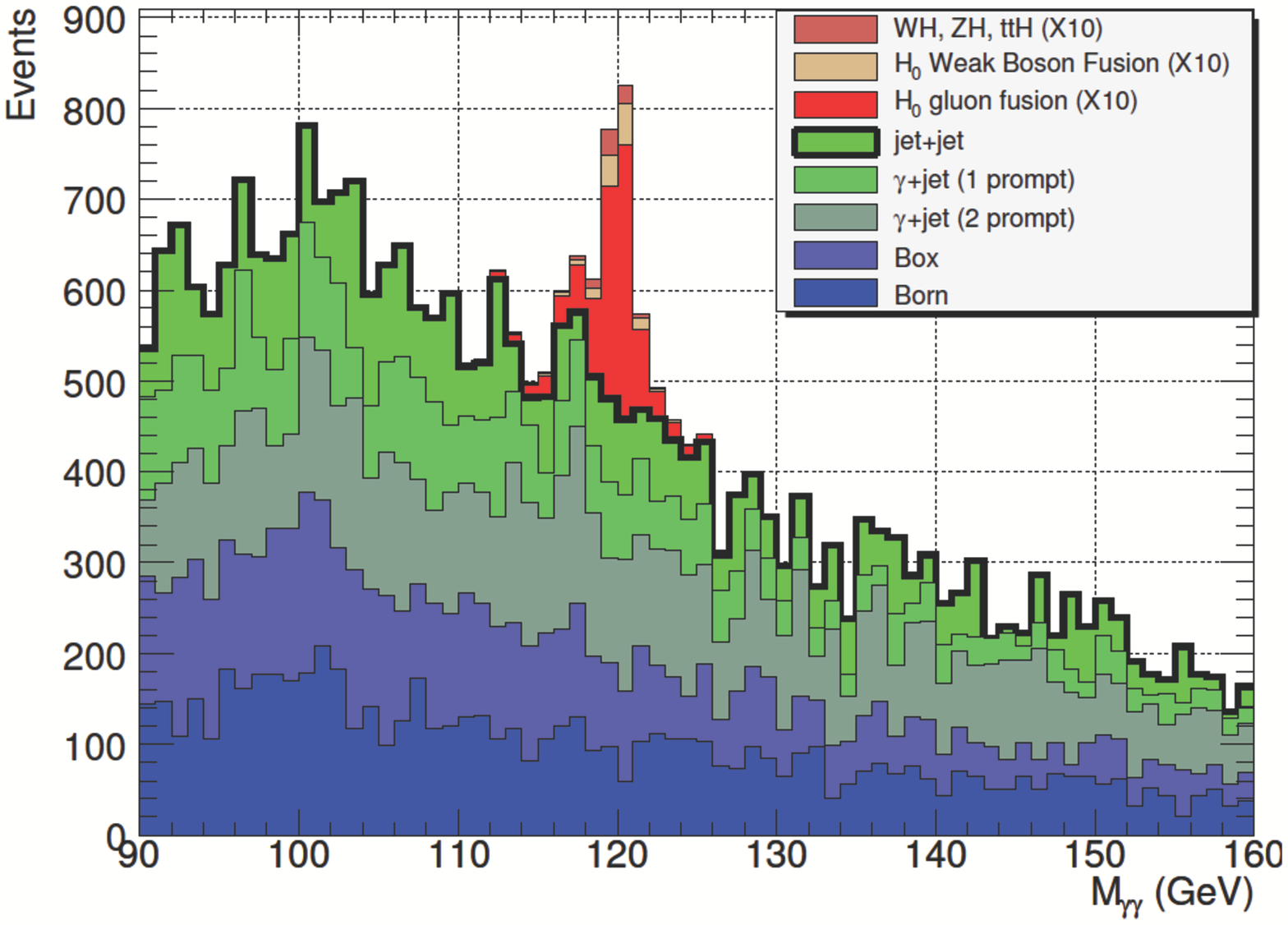}~
 \includegraphics[width=0.45\linewidth]{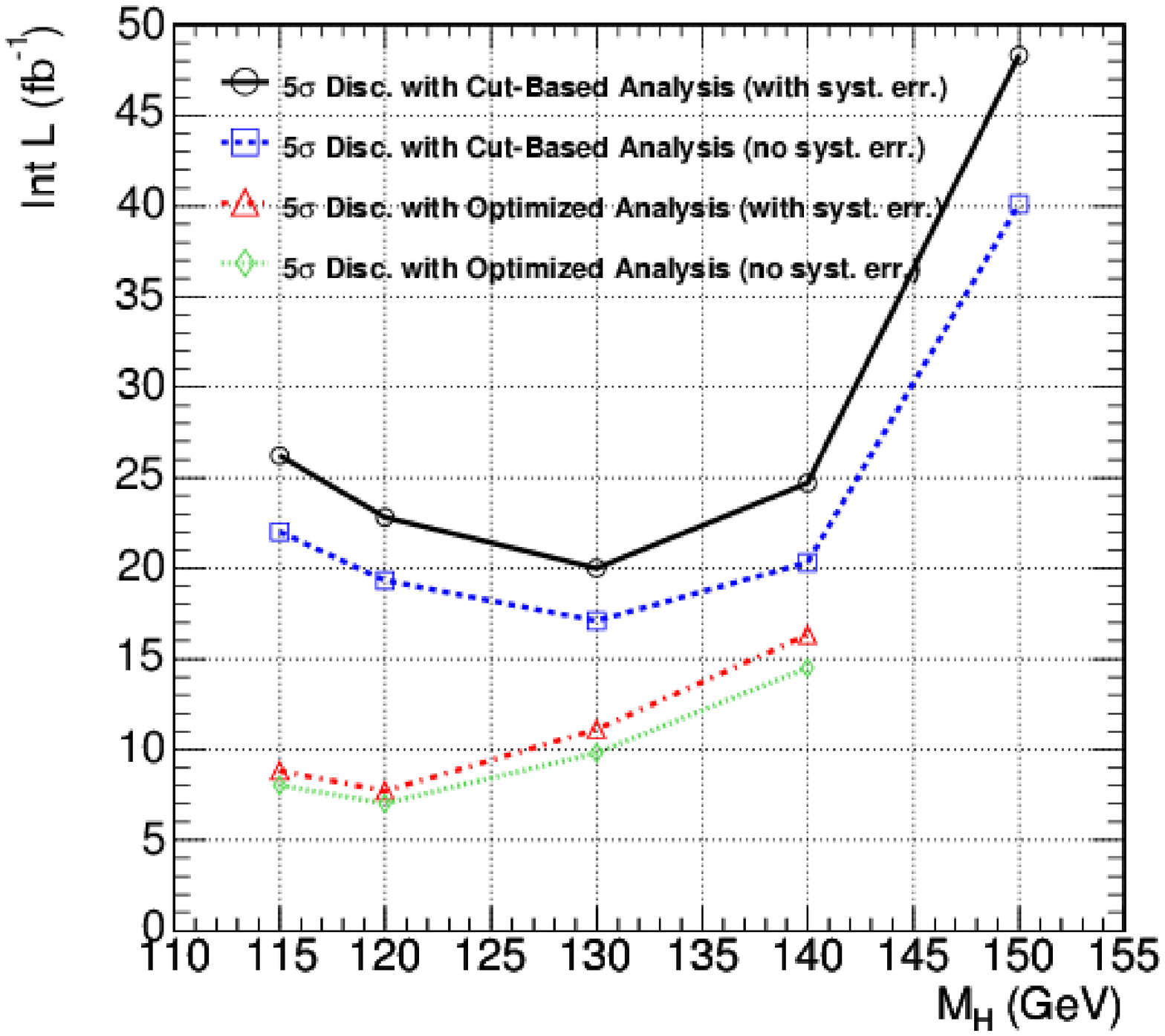}
\caption{Left: The diphoton mass distribution for each source for barrel events with kinematic neural net output
greater than~0.85. Events are normalized to an integrated luminosity of 7.7~fb$^{-1}$ and the signal 
($M_\mathrm{H}=120$~GeV) is scaled by a factor~10. Right: Integrated luminosity needed for a $5 \sigma$ 
discovery with the cut-based and optimized analyses.} 
\label{fig:2}
\end{figure*}

To improve the search sensitivity, the selected events are split into categories based on the compactness 
of photon showers and the pseudo-rapidity of the photon candidates. This allows us to take advantage of
better mass resolution when we expect it, yet still use all of the selected events. 
The confidence levels are then computed by using the Log Likelihood Ratio frequentistic method~\cite{cms:hgg}.
Figure~\ref{fig:2} shows that for $M_\mathrm{H} < 140$~GeV 
this ``cut-based'' approach should lead to a $5\sigma$ discovery with less than 30~fb$^{-1}$ of integrated
luminosity.  Moreover, if the Higgs boson will be discovered in this
channel, we will be able to measure its mass with a statistical precision of $0.1-0.2\%$, already with 
about 30~fb$^{-1}$.
On the other hand, approximately 5~fb$^{-1}$ are needed for a 95\% C.L. exclusion in the same mass range if the
 SM Higgs boson does not exist in that mass range.

A different search strategy, referred to as  the “optimized analysis,” consists in using
 a neural net to discriminate between the signal 
and background. Photon isolation and event kinematics variables are used as inputs. 
Signal significance is derived using the neural net output and reconstructed diphoton invariant mass. 
Figure~\ref{fig:2} shows the mass distribution after  a cut on the neural net output at 0.85, for
$M_\mathrm{H}=120$~GeV. The optimized analysis should lead to a significantly better discovery
potential. As shown in Figure~\ref{fig:2}, for  $M_\mathrm{H} < 140$~GeV the 
 $\mathrm{H} \rightarrow\gamma\gamma$  signal should be discovered with about 10~fb$^{-1}$~\cite{cms:hgg}. 

\subsection{Vector Boson Fusion Production Mechanism}

Establishing $qq \rightarrow \mathrm{H}qq$ production via vector-boson fusion (VBF) 
will be crucial for understanding the exact nature of the Higgs mechanism. 
Such events will be characterized by the presence of two 
back-to-back high-rapidity jets.  
Therefore, the forward jet tagging should result in a clean separation of the VBF signal from 
the gluon fusion signal and the majority of backgrounds. For  $M_\mathrm{H} < 140$~GeV, a 
 $5\sigma$ discovery of the  
$\mathrm{H} \rightarrow\gamma\gamma$ VBF signal 
can be achieved with about $50-100$~fb$^{-1}$ of integrated luminosity~\cite{cms:hggvbf}. 

\section{SUMMARY}
With a standard cut-based analysis with less than 30~fb$^{-1}$ of
 integrated luminosity we can discover the SM Higgs boson
with $5\sigma$ significance between the LEP lower limit and 140 GeV.
A significantly better discovery potential can be achieved using an optimized analysis, for which
only about 10~fb$^{-1}$ will be needed to establish the $\mathrm{H} \rightarrow\gamma\gamma$ signal.
The detector resolution for the reconstructed  Higgs boson mass profits from the excellent energy resolution 
of the CMS crystal calorimeter.

This paper provides only a brief sketch of the analysis methods and results. For more information,
the reader is invited to consult the provided references as well as the CMS Physics TDR. 

\begin{acknowledgments}
I  thank all my colleagues from the CMS collaboration. In particular,
I am grateful to Drs.~Yves Sirois and Chiara Mariotti for their help.
This work was supported in part by  the 
 U.S. Department of Energy Grant No.~DE-FG03-92-ER40701.

\end{acknowledgments}

\end{document}